\documentclass[12pt]{article}
\usepackage{amsmath}
\def\be{\begin{equation}}
\def\ee{\end{equation}}
\def\arr{\begin{array}{rll}}
\def\ea{\end{array}}
\def\bea{\begin{eqnarray}}
\def\eea{\end{eqnarray}}

\def\N2{$N{=}2$}

\def\>{\rangle}
\def\<{\langle}
\def\+{\dagger}
\def\={\ =\ }

\begin{document}
\renewcommand{\thefootnote}{\fnsymbol{footnote}}
\begin{titlepage}
\setcounter{page}{0}
\begin{flushright}
LMP-TPU--9/12  \\
\end{flushright}
\vskip 1cm
\begin{center}
{\LARGE\bf Dynamical realization of $l$--conformal }\\
\vskip 0.5cm
{\LARGE\bf Galilei algebra and oscillators}\\
\vskip 1cm
$
\textrm{\Large Anton Galajinsky and Ivan Masterov\ }
$
\vskip 0.7cm
{\it
Laboratory of Mathematical Physics, Tomsk Polytechnic University, \\
634050 Tomsk, Lenin Ave. 30, Russian Federation} \\
{E-mails: galajin@mph.phtd.tpu.ru, masterov@mph.phtd.tpu.ru}

\end{center}
\vskip 1cm
\begin{abstract} \noindent
The method of nonlinear realizations is applied to
the $l$--conformal Galilei algebra to construct a dynamical system
without higher derivative terms in the equations of motion. A configuration space of the model involves coordinates, which
parametrize particles in $d$ spatial dimensions, and
a conformal mode, which gives rise to an effective external field. It is shown that trajectories of the system can
be mapped into those of a set of decoupled oscillators in $d$ dimensions.
\end{abstract}


\vskip 0.5cm


\end{titlepage}

\renewcommand{\thefootnote}{\arabic{footnote}}
\setcounter{footnote}0

\noindent
{\bf 1. Introduction}\\

\noindent

It has long been realized that Galilei algebra can be extended by conformal generators in more ways than one \cite{hp,hen1,nor}.
In general, a conformal extension of the Galilei algebra is parametrized by a positive
half integer $l$, which justifies the term $l$--conformal Galilei algebra\footnote{In modern literature the reciprocal of $l=\frac{N}{2}$, where $N=1,2,\dots$, is called the rational dynamical exponent.
The corresponding
algebra is sometimes referred to as
the conformal Galilei algebra with rational dynamical exponent or $N$--Galilean conformal algebra.
In this work we stick to the terminology
originally adopted in \cite{nor}.}. The instance of
$l=\frac 12$, known in the literature as the Schr\"odinger algebra, has been the focus of most studies (for a review see e.g. \cite{PH}).
Motivated by current investigation of the non--relativistic version of the AdS/CFT correspondence, conformal
Galilei algebras with $l>\frac 12$ have recently attracted considerable attention \cite{lsz1}--\cite{ai}.

In general, $(2l+1)$ vector generators enter the $l$--conformal Galilei algebra. Apart from the spatial
translations and Galilei boosts, one reveals accelerations for $l>\frac 12$.
When constructing a dynamical realization, the generators in the algebra are linked to constants of the motion, which facilitate
solving the equations of motion. Because the number of functionally independent constants of the motion needed to integrate
a differential equation correlates with its order, dynamical realizations of the $l$--conformal Galilei algebra in general involve higher
derivative terms (see e.g. \cite{lsz1,dh1,gk,agm}).

A dynamical realization of $l=1$ conformal Galilei algebra, which is free from higher derivative terms, has been recently constructed in
\cite{fil} within the method of nonlinear realizations \cite{cwz,ccwz}. It will be demonstrated below that the reason why one can accommodate
$l=1$ conformal Galilei symmetry in second order differential equations is that the generator of accelerations is functionally dependent
on those related to the spatial
translations and Galilei boosts. At this point it is worthwhile drawing an analogy with the conformal particle in one dimension
\cite{dff}. The model holds invariant under the conformal group $SO(2,1)$, which involves three generators corresponding to time translations,
dilatations and special conformal transformations. However, as far as the dynamical realization \cite{dff} is concerned,
the latter generator proves to be functionally dependent on the former. To put it in other words, constants of the motion related to the time translations
and dilatations are sufficient to solve the equation of motion.

The purpose of this work is to generalize the analysis in \cite{fil} to the case of arbitrary $l$, i.e. to construct a dynamical
realization of the $l$--conformal Galilei algebra in terms of second order differential equations.

The work is organized as follows. In the next section, the method of nonlinear realizations is applied to the $l$--conformal Galilei algebra.
In particular, we choose an appropriate coset space, derive transformation laws for coordinates which parametrize it,
construct the left--invariant
Maurer--Cartan one--forms and impose constraints on them, which determine the equations of motion.
Sect. 3 contains two explicit examples of
$l=1$ and $l=2$, which facilitate the analysis of general $l$ in Sect. 4. It is also shown in Sect. 4 that for half--integer $l$ there is more than one
way to construct a dynamical system. An alternative realization of $l=\frac 32$ is considered in Sect. 5. We conclude with the discussion
of possible further developments in Sect. 6.

\vspace{0.5cm}

\noindent
{\bf 2. Dynamical realization of the $l$--conformal Galilei algebra}\\

\noindent

The $l$--conformal Galilei algebra includes the generators of time translations, dilatations, special conformal transformations, spatial rotations,
spatial translations, Galilei boosts
and accelerations. Denoting the generators by $(H,D,K, M_{ij}, C^{(n)}_i)$, respectively, where $i=1,\dots,d$ is a spatial index and $n=0,1,\dots, 2l$,
one has the structure relations \cite{nor}
\begin{align}\label{algebra}
&
[H,D]=i H, &&  [H,C^{(n)}_i]=i n C^{(n-1)}_i,
\\[2pt]
&
[H,K]=2 i D, && [D,K]=i K,
\nonumber\\[2pt]
&
[D,C^{(n)}_i]=i (n-l) C^{(n)}_i, && [K,C^{(n)}_i]=i (n-2l) C^{(n+1)}_i,
\nonumber\\[2pt]
&
[M_{ij},C^{(n)}_k]=-i (\delta_{ik} C^{(n)}_j-\delta_{jk} C^{(n)}_i), && [M_{ij},M_{kl}]=-i(\delta_{ik} M_{jl}+\delta_{jl} M_{ik}-
\delta_{il} M_{jk}-\delta_{jk} M_{il}).
\nonumber
\end{align}
Note that $(H,D,K)$ form $so(2,1)$ subalgebra, which is the conformal algebra in one dimension.
The instances of $n=0$ and $n=1$ in $C^{(n)}_i$ correspond to the spatial translations and Galilei boosts.
Higher values of $n$ are linked to the accelerations.

In order to construct second order differential equations, which holds invariant under the $l$--conformal Galilei group,
we choose to apply the method of nonlinear realizations \cite{cwz,ccwz,io} to the algebra (\ref{algebra}). Note that the instance of $l=1$
has been previously studied in
\cite{fil}.

As the first step, one considers the coset space\footnote{As usual, summation over repeated indices is understood.}
\be
\tilde G=e^{itH} e^{izK} e^{iuD} e^{i x^{(n)}_i C^{(n)}_i} \times SO(d)
\ee
parametrized by the coordinates $(t,z,u, x^{(n)}_i)$.
Left multiplication by a group element
$g=e^{iaH} e^{ibK} e^{icD} e^{i \lambda^{(n)}_i C^{(n)}_i} e^{\frac{i}{2} \omega_{ij} M_{ij}}$
determines the action of the group on the coset space.
Taking into account the Baker--Campbell--Hausdorff formula
\be\label{ser}
e^{iA}~ T~ e^{-iA}=T+\sum_{n=1}^\infty\frac{i^n}{n!}
\underbrace{[A,[A, \dots [A,T] \dots]}_{n~\rm times},
\ee
one derives the infinitesimal coordinate transformations
\bea\label{transf}
&&
\delta t=a+b t^2+ct, \qquad \delta z=b(1-2tz)-cz, \qquad \delta u=c+2bt, \qquad \delta x^{(n)}_i=-\omega_{ij} x^{(n)}_j,
\nonumber\\[2pt]
&&
\delta x^{(n)}_i=e^{u(n-l)} \sum_{s=0}^n \sum_{m=s}^{2l}\frac{{(-1)}^{n-s} m! (2l-s)!}{s! (m-s)! (n-s)! (2l-n)!} t^{m-s} z^{n-s} \lambda^{(m)}_i,
\eea
where $a,b,c,\lambda^{(n)}_i$ and $\omega_{ij}$
are infinitesimal parameters corresponding to time translations, special conformal transformations, dilatations, vector generators in the algebra 
and spatial rotations, respectively. When obtaining (\ref{transf}), the identity
\be
\sum_{p=0}^{2l} \sum_{m=0}^p k_{pm}=\sum_{m=0}^{2l} \sum_{p=m}^{2l} k_{pm},
\ee
where $k_{pm}$ is an arbitrary matrix, proves to be helpful.

As the next step, one considers the subgroup $G=e^{itH} e^{izK} e^{iuD} e^{i x^{(n)}_i C^{(n)}_i}$ and constructs the left--invariant
Maurer--Cartan one--forms
\be
G^{-1} d G=i(\omega_H H+\omega_K K+\omega_D D+\omega^{(n)}_i C^{(n)}_i),
\ee
where we denoted
\bea\label{MC}
&&
\omega_H=e^{-u} dt, \qquad \omega_K=e^u (z^2 dt+dz), \qquad \omega_D=d u-2z dt,
\nonumber\\[2pt]
&&
\omega^{(n)}_i=d x^{(n)}_i-(n-l) x^{(n)}_i \omega_D -(n+1)  x^{(n+1)}_i \omega_H-(n-2l-1)  x^{(n-1)}_i \omega_K.
\eea
In the last line it is assumed that $x^{(-1)}_i=x^{(2l+1)}_i=0$. By construction the one--forms (\ref{MC}) are invariant under all the transformations in
(\ref{transf}) but for rotations with respect to which $\omega^{(n)}_i$ is transformed as a vector.

The one--forms (\ref{MC}) are the clue to a dynamical realization. Setting some of them to zero
one can either reduce the number of degrees of freedom via algebraic equations or obtain reasonable dynamical equations of motion, which are automatically
invariant under the action of a given group \cite{io}. Choosing the constraints requires guesswork, however.
For the conformal subalgebra $so(2,1)$ we
follow the recipe in \cite{ikl} and impose the restrictions
\be\label{const}
\omega_D=0, \qquad \gamma^{-1} \omega_K-\gamma \omega_H=0,
\ee
where $\gamma$ is an arbitrary (coupling) constant. Taking $t$ to be the temporal coordinate and introducing the new variable
\be\label{ro}
\rho=e^{\frac{u}{2}},
\ee
from (\ref{const}) one gets
\be\label{z}
z=\frac{\dot\rho}{\rho}, \qquad \ddot\rho=\frac{\gamma^2}{\rho^3},
\ee
where the overdot denotes the derivative with respect to time.
Thus, given the constraints (\ref{const}), the variable $z$ is not independent and can be discarded, while $\rho$ describes the conformal particle in one
dimension \cite{dff,ikl}.

For the variables $x^{(n)}_i$ we choose the following constraint
\be
\omega^{(n)}_i=0,
\ee
which, in view of (\ref{const}), yields
\be\label{eqx}
\rho^2 \dot x^{(n)}_i=(n+1) x^{(n+1)}_i- (2l-n+1) \gamma^2 x^{(n-1)}_i.
\ee
It is to be remembered that $x^{(-1)}_i=x^{(2l+1)}_i=0$ by definition. A dynamical realization of the $l$--conformal Galilei algebra is thus encoded in
the master equations (\ref{eqx}). In the next section we consider in detail the examples of $l=1$ and $l=2$, which facilitate the
analysis of arbitrary $l$ in Sec. 4.

\vspace{0.5cm}

\noindent
{\bf 3. Examples}\\


\noindent
{\it 3.1. The case of $l=1$}\\

\noindent

For $l=1$ the equations (\ref{eqx}) amount to
\be\label{l1}
\rho^2 \dot x^{(0)}_i=x^{(1)}_i, \qquad \rho^2 \dot x^{(1)}_i=2 x^{(2)}_i-2 \gamma^2 x^{(0)}_i, \qquad \rho^2 \dot x^{(2)}_i=-\gamma^2 x^{(1)}_i.
\ee
If one treated (\ref{l1}) literally, one would eliminate $x^{(1)}_i$ and $x^{(2)}_i$ via the first two algebraic relations and obtain the third
order differential equation for $x^{(0)}_i$ from the rightmost restriction in (\ref{l1})
\be
\rho^2 \frac{d}{dt} \left( \rho^2 \frac{d}{dt} \left(\rho^2 \frac{d}{dt} x^{(0)}_i \right) \right) +4 \gamma^2 \rho^2 \frac{d}{dt} x^{(0)}_i=0.
\ee
This approach is analogous to the higher derivative realizations of $l=1$ conformal Galilei algebra considered recently in \cite{gk,agm}.

An alternative possibility is to get rid of $x^{(1)}_i$ via the first equation in (\ref{l1}), bring the remaining equations to the form
\be\label{eqchi}
\rho^2 \frac{d}{dt} \left(\rho^2 \frac{d}{dt} \chi_i \right)+4 \gamma^2 \chi_i=0, \qquad  \rho^2 \frac{d}{dt} \zeta_i=0,
\ee
where
\be\label{chi}
\chi_i=\gamma x^{(0)}_i-\gamma^{-1} x^{(2)}_i, \qquad \zeta_i=\gamma x^{(0)}_i+\gamma^{-1} x^{(2)}_i,
\ee
and discard $\zeta_i$ as obeying the first order differential equation. Following this road one is left with two second order
differential equations
\be\label{L1}
\ddot\rho=\frac{\gamma^2}{\rho^3}, \qquad \rho^2 \frac{d}{dt} \left(\rho^2 \frac{d}{dt} \chi_i \right)+4 \gamma^2 \chi_i=0,
\ee
which provide a dynamical realization of $l=1$ conformal Galilei algebra. Note that the variables $\rho$ and $\chi_i$ are separated in (\ref{L1}).
In particular, one can solve the differential equation for the conformal mode $\rho$ and substitute its solution to the equation for $\chi_i$.
The latter will then describe a particle in $d$ spatial dimensions moving in an external field.

Let us discuss in more detail the way in which $l=1$ conformal Galilei symmetry is realized in (\ref{L1}) and
how it facilitates integration of the equations of motion. As was shown above, the construction of a dynamical realization involves a passage from
the coset coordinates $(t,\rho,\chi_i)$ to the fields $(\rho(t),\chi_i (t))$ obeying the equations of motion (\ref{L1}).
On the space of fields $l=1$ conformal Galilei group acts as follows:
\be
\rho' (t')=\rho(t)+\delta \rho, \qquad \chi'_i (t')=\chi_i(t)+\delta \chi_i,
\ee
where $\delta \rho$ and $\delta \chi_i$ are inherited from the coset transformations (\ref{transf}). Taking into account that $\rho$ is related to
$u$ in (\ref{ro}), $\chi_i$ is linked to $x^{(0)}_i, x^{(2)}_i$ in (\ref{chi}) and $z$ is fixed in (\ref{z}), one derives transformation laws for
the form of the fields
\bea\label{form}
&&
\rho'(t)=\rho(t)+\frac{1}{2} (c+2bt) \rho(t)-(a+b t^2+ct) \dot\rho(t),
\nonumber\\[2pt]
&&
\chi'_i (t)=\chi_i (t)-(a+bt^2+ct)\dot\chi_i(t)+\left(\frac{\gamma}{\rho^2}-\frac{\dot\rho^2}{\gamma}\right) \lambda^{(0)}_i+
\left(t \left(\frac{\gamma}{\rho^2}-\frac{\dot\rho^2}{\gamma}\right)+\frac{\rho \dot\rho}{\gamma} \right) \lambda^{(1)}_i
\nonumber\\[2pt]
&& \qquad \quad
+\left(t^2 \left(\frac{\gamma}{\rho^2}-\frac{\dot\rho^2}{\gamma}\right)+\frac{2t \rho \dot\rho}{\gamma} -\frac{\rho^2}{\gamma}\right) \lambda^{(2)}_i.
\eea
It is straightforward to verify that the equations of motion (\ref{L1}) are invariant under the infinitesimal transformations (\ref{form}).
Furthermore, considering the variation of the form of the fields $\delta \rho(t)=\rho'(t)-\rho(t)$, $\delta \chi_i(t)=\chi'_i(t)-\chi(t)$ and computing the
commutator\footnote{When evaluating the commutator, it is to be understood that the variation acts on the form of a field only and
does not affect the temporal coordinate $t$, which appears explicitly on the right hand side of (\ref{form}).} $[\delta_1,\delta_2]$,
one can reproduce the algebra (\ref{algebra}) with $l=1$.

Now let us link the symmetry transformations (\ref{form}) to constants of the motion. Although the system of equations (\ref{L1}) is not Lagrangian, the fact that the variables are separated
allows one to deal with effective Lagrangians and to build constants of the motion via the Noether theorem. Consider the first equation in (\ref{L1}).
The action functional corresponding to it reads \cite{dff}
\be
S_{\rho}=\int dt \left( \dot\rho^2-\frac{\gamma^2}{\rho^2} \right),
\ee
which holds invariant under the conformal transformations considered above
\be
t'=t+a+b t^2+ct, \qquad \rho'(t')=\rho(t)+\frac{1}{2} (c+2bt) \rho(t).
\ee
The Noether theorem then yields constants of the motion. To keep track of the $l$--conformal Galilei algebra here and in what follows we designate
constants of the motion by the same letters as in
(\ref{algebra}) but in a calligraphic style\footnote{
Lagrangian symmetry transformations which we consider in this work are of the form
$t'=t+\Delta t (t)$, $x'_i (t')=x_i(t)+\Delta x_i(t,x(t))$.
If the action functional $S=\int dt \mathcal{L}(x,\dot x)$ holds invariant under the transformation up to a total derivative,
i.e. $\delta S=\int dt \Big( \frac{d F}{dt}\Big)$, then
the conserved quantity is derived from the expression
$\Delta x_i \frac{\partial \mathcal{L}}{\partial \dot x_i}-
\Delta t \Big( \dot x_i \frac{\partial \mathcal{L}}{\partial \dot x_i}-\mathcal{L} \Big)-F$
by discarding the parameter of the transformation.}
\be\label{hdk}
\mathcal{H}=\dot\rho^2+\frac{\gamma^2}{\rho^2}, \qquad \mathcal{D}=\rho \dot\rho-t \mathcal{H}, \qquad \mathcal{K}=t^2  \mathcal{H}-2t \rho \dot\rho+\rho^2.
\ee
$\mathcal{H}$ and $\mathcal{D}$ allow one to fix $\rho(t)$
\be\label{sol}
\rho(t)=\sqrt{\frac{{(\mathcal{D}+t\mathcal{H})}^2+\gamma^2}{\mathcal{H}}},
\ee
where for definiteness we have chosen a positive root of a quadratic algebraic equation, which determines $\rho$.
Note that the last constant of the motion in (\ref{hdk}) is functionally dependent on the others
\be
\mathcal{K}=\frac{\mathcal{D}^2+\gamma^2}{\mathcal{H} },
\ee
as it should be the case because one needs only two constants of the motion in order to integrate a second order differential equation.

Having fixed $\rho$, one can consider an effective action for the second equation in (\ref{L1})
\be\label{ea}
S_{\chi}=\int dt \left( \rho^2 \dot\chi_i  \dot\chi_i -4 \gamma^2 \frac{ \chi_i \chi_i}{\rho^2} \right),
\ee
where $\rho$ is to be treated as a background field obeying the first equation in (\ref{L1}).
This action is invariant under the transformations
\bea\label{trl1}
&&
t'=t, \qquad
\chi'_i (t')=\chi_i (t)+\left(\frac{\gamma}{\rho^2}-\frac{\dot\rho^2}{\gamma}\right) \lambda^{(0)}_i+
\left(t \left(\frac{\gamma}{\rho^2}-\frac{\dot\rho^2}{\gamma}\right)+\frac{\rho \dot\rho}{\gamma} \right) \lambda^{(1)}_i
\nonumber\\[2pt]
&&
\qquad \quad \qquad \qquad \quad
+\left(t^2 \left(\frac{\gamma}{\rho^2}-\frac{\dot\rho^2}{\gamma}\right)+\frac{2t \rho \dot\rho}{\gamma} -\frac{\rho^2}{\gamma}\right) \lambda^{(2)}_i
-\omega_{ij} \chi_j (t),
\eea
which result in constants of the motion
\bea\label{int}
&&
{\mathcal{C}}^{(0)}_i=\rho^2 \dot\chi_i \left(\frac{\gamma}{\rho^2}-\frac{\dot\rho^2}{\gamma} \right) +4 \gamma \chi_i \frac{\dot\rho}{\rho}, \qquad
\qquad \quad \quad {\mathcal{C}}^{(1)}_i=t {\mathcal{C}}^{(0)}_i+\dot\chi_i \frac{\rho^3 \dot\rho}{\gamma}-2\gamma\chi_i,
\nonumber\\[2pt]
&&
{\mathcal{C}}^{(2)}_i=t^2 {\mathcal{C}}^{(0)}_i+2t \left( \dot\chi_i \frac{\rho^3 \dot\rho}{\gamma}-2\gamma\chi_i\right)- \dot\chi_i \frac{\rho^4}{\gamma},
\qquad \mathcal{M}_{ij}=\rho^2 (\chi_i \dot\chi_j-\chi_j \dot\chi_i).
\eea
The first two can be used to determine the evolution of $\chi_i(t)$ with time
\be\label{sol1}
\chi_i(t)=\alpha_i \cos{(2\gamma s(t))}+\beta_i \sin{(2\gamma s(t))},
\ee
where $\alpha_i$ and $\beta_i$ are constants of integration\footnote{The constants
$\alpha_i$ and $\beta_i$ are related to ${\mathcal{C}}^{(0)}_i$, ${\mathcal{C}}^{(1)}_i$, $\mathcal{H}$ and $\mathcal{D}$ as follows:
$\alpha_i=-\frac{1}{2 \gamma} \left({\mathcal{C}}^{(1)}_i+\frac{\mathcal{D}}{\mathcal{H}}{\mathcal{C}}^{(0)}_i\right)$,
$\beta_i=\frac{{\mathcal{C}}^{(0)}_i}{2\mathcal{H}}$.} and
$s(t)$ is a subsidiary function
\be\label{s}
s(t)=\frac{1}{\gamma} \arctan{\left(\frac{\mathcal{D}+t \mathcal{H}}{\gamma}\right)}, \qquad \dot s(t)=\frac{1}{\rho^2}.
\ee
Some useful formulae relating $t$, $\rho$, $\dot\rho$ and $s$ read
\be\label{t}
t=\frac{\gamma \tan{(\gamma s)}-\mathcal{D}}{\mathcal{H}}, \qquad \rho=\frac{\gamma}{\sqrt{\mathcal{H}} \cos{(\gamma s)}}, \qquad \dot\rho=\sqrt{\mathcal{H}} \sin{(\gamma s)}.
\ee
These will be extensively used below.

Like $\mathcal{K}$ considered above, ${\mathcal{C}}^{(2)}_i$ proves to
be functionally dependent
\be
{\mathcal{C}}^{(2)}_i=-\left(\frac{{\mathcal{D}}^2+\gamma^2}{{\mathcal{H}}^2} \right) {\mathcal{C}}^{(0)}_i-\frac{2 \mathcal{D}}{\mathcal{H}}
{\mathcal{C}}^{(1)}_i,
\ee
which correlates well with the fact that two independent constants of the motion are enough to integrate a second order differential equation.
Note that the redundancy of the transformation with the parameter $\lambda^{(2)}_i$ can be also revealed by looking at the generators in
(\ref{trl1}). Given the explicit form of $\rho$ in (\ref{sol}), one readily finds the identity
\be
\frac{{\mathcal{D}}^2+\gamma^2}{{\mathcal{H}}^2} \left(\frac{\gamma}{\rho^2}-\frac{\dot\rho^2}{\gamma}\right)
+\frac{2 \mathcal{D}}{\mathcal{H}}\left(t \left(\frac{\gamma}{\rho^2}-\frac{\dot\rho^2}{\gamma}\right)+\frac{\rho \dot\rho}{\gamma} \right)
+t^2 \left(\frac{\gamma}{\rho^2}-\frac{\dot\rho^2}{\gamma}\right)+\frac{2t \rho \dot\rho}{\gamma} -\frac{\rho^2}{\gamma}=0.
\ee

To summarize, for the dynamical system governed by the second order differential equations (\ref{L1}) $l=1$ conformal Galilei symmetry results in the set of constants of the motion
(\ref{hdk}), (\ref{int}), which allow one to derive the general solution of the equations of motion in a rather efficient way.
As a matter of fact, the transformations generated by $H$, $D$ and $K$ are essential for the conformal mode $\rho(t)$, while the vector generators 
$C^{(n)}_i$ play a central role for $\chi_i(t)$. It is noteworthy that the conformal mode not only provides a source of an effective external field 
for $\chi_i$ in (\ref{L1}), but it is also a principal ingredient in constructing the vector transformations  (\ref{trl1}) acting on $\chi_i(t)$.

Before concluding this section, it is worth mentioning that (\ref{s}) and (\ref{t}) link the second equation in
(\ref{L1}) to an ordinary harmonic oscillator in $d$ dimensions. In terms of the variable $s$ the second equation in
(\ref{L1}) reads
\be\label{osc}
\frac{d^2 \chi_i}{ds^2} +4 \gamma^2 \chi_i=0,
\ee
which also elucidates the form of the solution (\ref{sol1}). Thus, the shape of an orbit traced by a particle parametrized by  
$\chi_i(t)$ is analogous to that of the oscillator (\ref{osc}). However, in contrast to (\ref{osc}) the orbit of $\chi_i(t)$ is not closed 
(see (\ref{s}) above).

Note that the relation between $t$ and $s$ in (\ref{t}) resembles Niederer's
transformation \cite{nied1}, which is known to relate the $l$--conformal Galilei algebra to its Newton-Hooke counterpart \cite{gm}.
In particular, being rewritten in terms of the variable $s$, transformations with the parameters $\lambda^{(0)}_i$ and $\lambda^{(1)}_i$
precisely reproduce the spatial translations and Galilei boosts realized in the harmonic oscillator (see e.g. \cite{gala}).

\vspace{0.5cm}

\noindent
{\it 3.2. The case of $l=2$}\\

\noindent

For $l=2$ the master equations (\ref{eqx}) read
\begin{align}\label{l2}
&
\rho^2 \dot x^{(0)}_i=x^{(1)}_i, && \rho^2 \dot x^{(1)}_i=2 x^{(2)}_i-4 \gamma^2 x^{(0)}_i, && \rho^2 \dot x^{(2)}_i=3 x^{(3)}_i-3 \gamma^2 x^{(1)}_i,
\nonumber\\[2pt]
&
\rho^2 \dot x^{(3)}_i=4 x^{(4)}_i-2 \gamma^2 x^{(2)}_i, && \rho^2 \dot x^{(4)}_i=-\gamma^2 x^{(3)}_i. &&
\end{align}
One can use the first and the last relations in (\ref{l2}) to remove $x^{(1)}_i$ and $x^{(3)}_i$. The remaining equations decouple after one introduces
the new variables
\bea\label{nvar}
&&
\zeta_i=3\gamma^2 x^{(0)}_i+x^{(2)}_i+\frac{3}{\gamma^2} x^{(4)}_i,
\nonumber\\[2pt]
&&
\chi_i=\gamma^2 x^{(0)}_i-\frac{1}{\gamma^2} x^{(4)}_i,
\nonumber\\[2pt]
&&
\xi_i=\gamma^2 x^{(0)}_i-x^{(2)}_i+\frac{1}{\gamma^2} x^{(4)}_i.
\eea
These give
\bea\label{L2}
\rho^2 \frac{d}{dt} \left(\rho^2 \frac{d}{dt} \chi_i \right)+4 \gamma^2 \chi_i=0, \qquad
\rho^2 \frac{d}{dt} \left(\rho^2 \frac{d}{dt} \xi_i \right)+16 \gamma^2 \xi_i=0,
\eea
along with $\rho^2 \frac{d}{dt} \zeta_i=0$ for $\zeta_i$.
It is understood that (\ref{L2}) is
accompanied with $\ddot\rho=\frac{\gamma^2}{\rho^3}$.
In what follows, we disregard $\zeta_i$ as it obeys the first order differential equation.

Note that in (\ref{L2}) we encounter the same equation, which appeared above in (\ref{L1}) for $l=1$. Thus, the first equation in
(\ref{L2}) may accommodate both $l=1$ and $l=2$ conformal Galilei symmetries. The reason why this is possible will be explained in the next section.

Consider the first equation in (\ref{L2}).
The transformation law of the field $\chi_i(t)$ is deduced from (\ref{transf}), (\ref{ro}) and the first relation in (\ref{z})
\bea\label{tl2}
&&
t'=t, \qquad \chi'_i(t')=\chi_i(t)+\sum_{n=0}^4 v^{(n)} \lambda^{(n)},
\\[2pt]
&&
v^{(0)}=\frac{\gamma^2}{\rho^4}-\frac{\dot\rho^4}{\gamma^2}, \qquad v^{(1)}=t v^{(0)}+\frac{\rho \dot\rho^3}{\gamma^2}, \qquad v^{(2)}=t^2 v^{(0)}+2 t \frac{\rho \dot\rho^3}{\gamma^2}
-\frac{\rho^2 \dot\rho^2}{\gamma^2},
\nonumber\\[2pt]
&&
v^{(3)}=t^3 v^{(0)}+3 t^2 \frac{\rho \dot\rho^3}{\gamma^2}
-3 t\frac{\rho^2 \dot\rho^2}{\gamma^2}+\frac{\rho^3 \dot\rho}{\gamma^2}, \qquad v^{(4)}=t^4 v^{(0)}+4 t^3 \frac{\rho \dot\rho^3}{\gamma^2}
-6 t^2\frac{\rho^2 \dot\rho^2}{\gamma^2}+4 t \frac{\rho^3 \dot\rho}{\gamma^2}-\frac{\rho^4}{\gamma^2}.
\nonumber
\eea
Taking into account the explicit form of $\rho$ in (\ref{sol}), one reveals the identities
\bea
&&
v^{(2)}+\frac{{\mathcal{D}}^2}{{\mathcal{H}}^2} v^{(0)}+\frac{2 \mathcal{D}}{\mathcal{H}} v^{(1)}=0, \qquad
v^{(3)}-\frac{\mathcal{D} (2{\mathcal{D}}^2+\gamma^2)}{{\mathcal{H}}^3} v^{(0)}-\frac{(3 {\mathcal{D}}^2+\gamma^2)}{{\mathcal{H}}^2} v^{(1)}=0,
\nonumber\\[2pt]
&&
v^{(4)}+\frac{(3{\mathcal{D}}^4+4 {\mathcal{D}}^2 \gamma^2+\gamma^4)}{{\mathcal{H}}^4} v^{(0)}+\frac{4 \mathcal{D} ({\mathcal{D}}^2+\gamma^2)}{{\mathcal{H}}^3}  v^{(1)}=0,
\eea
which imply that constants of the motion corresponding to the parameters $\lambda^{(2)}_i$, $\lambda^{(3)}_i$, $\lambda^{(4)}_i$ can be expressed
via those related to $\lambda^{(0)}_i$, $\lambda^{(1)}_i$ and, as thus, they can be discarded.

Then one can construct the effective action, which reads as in (\ref{ea}), verify that it holds invariant under
the transformations (\ref{tl2}) with vector parameters $\lambda^{(0)}_i$, $\lambda^{(1)}_i$,  and build constants of the motion by applying the Noether theorem
\bea
&&
\mathcal{C}^{(0)}_i=\rho^2 \dot\chi_i \left(\frac{\gamma^2}{\rho^4}-\frac{\dot\rho^4}{\gamma^2}\right) +
4 \chi_i \frac{\dot\rho}{\rho} \left(\frac{\gamma^2}{\rho^2}+\dot\rho^2 \right),
\nonumber\\[2pt]
&&
\mathcal{C}^{(1)}_i=t \mathcal{C}^{(0)}_i+\dot\chi_i \frac{\rho^3 \dot\rho^3}{\gamma^2}-
\chi_i \left(\frac{\gamma^2}{\rho^2}+3 \dot\rho^2 \right).
\eea
These allow one to fix $\chi_i$
\be\label{Chi}
\chi_i(t)=\alpha_i \cos{(2\gamma s(t))}+\beta_i \sin{(2\gamma s(t))},
\ee
where $\alpha_i$ and $\beta_i$ are constants of integration related to $\mathcal{C}^{(0)}_i$, $\mathcal{C}^{(1)}_i$ and $s(t)$ is the subsidiary
function introduced in (\ref{s}). When deriving (\ref{Chi}), the formulae (\ref{t}) proved to be helpful.

The second equation in (\ref{L2}) can be treated likewise. The transformation law for the field $\xi_i(t)$ reads
\bea\label{trl2}
&&
t'=t, \qquad \xi'_i(t')=\xi_i(t)+\sum_{n=0}^4 u^{(n)} \lambda^{(n)},
\nonumber\\[2pt]
&&
u^{(0)}=\frac{\gamma^2}{\rho^4}+\frac{\dot\rho^4}{\gamma^2}-\frac{6 \dot\rho^2}{\rho^2},
\qquad u^{(1)}=t u^{(0)}-\frac{\rho \dot\rho^3}{\gamma^2}+\frac{3\dot\rho}{\rho},
\nonumber\\[2pt]
&&
u^{(2)}=t^2 u^{(0)}-2 t \left(\frac{\rho \dot\rho^3}{\gamma^2} -\frac{3\dot\rho}{\rho}\right)
+\frac{\rho^2 \dot\rho^2}{\gamma^2}-1,
\nonumber\\[2pt]
&&
u^{(3)}=t^3 u^{(0)}-3 t^2 \left( \frac{\rho \dot\rho^3}{\gamma^2} -\frac{3\dot\rho}{\rho}\right)
+3 t\left( \frac{\rho^2 \dot\rho^2}{\gamma^2} -1\right)  -\frac{\rho^3 \dot\rho}{\gamma^2},
\nonumber\\[2pt]
&&
u^{(4)}=t^4 u^{(0)}-4 t^3 \left( \frac{\rho \dot\rho^3}{\gamma^2}-\frac{3\dot\rho}{\rho} \right)
+6 t^2 \left( \frac{\rho^2 \dot\rho^2}{\gamma^2} -1\right) -4 t \frac{\rho^3 \dot\rho}{\gamma^2}+\frac{\rho^4}{\gamma^2}.
\eea
The functions $u^{(2)}$, $u^{(3)}$, $u^{(4)}$ prove to be linearly dependent on $u^{(0)}$ and $u^{(1)}$
\bea
&&
u^{(2)}+\frac{({\mathcal{D}}^2+\gamma^2)}{{\mathcal{H}}^2} u^{(0)}+\frac{2 \mathcal{D}}{\mathcal{H}} u^{(1)}=0, \qquad
u^{(3)}-\frac{2\mathcal{D} ({\mathcal{D}}^2+\gamma^2)}{{\mathcal{H}}^3} u^{(0)}-\frac{(3 {\mathcal{D}}^2-\gamma^2)}{{\mathcal{H}}^2} u^{(1)}=0,
\nonumber\\[2pt]
&&
u^{(4)}+\frac{(3{\mathcal{D}}^4+2 {\mathcal{D}}^2 \gamma^2-\gamma^4)}{{\mathcal{H}}^4} u^{(0)}
+\frac{4 \mathcal{D} ({\mathcal{D}}^2-\gamma^2)}{{\mathcal{H}}^3}  u^{(1)}=0,
\eea
and, hence, the transformations with the parameters $\lambda^{(2)}_i$, $\lambda^{(3)}_i$, $\lambda^{(4)}_i$ can be disregarded.

The effective action, which yields the second equation in (\ref{L2})
\be
S=\int dt \left( \rho^2 \dot\xi_i  \dot\xi_i -16 \gamma^2 \frac{ \xi_i \xi_i}{\rho^2} \right),
\ee
proves to be invariant under the transformations (\ref{trl2}),
of which only those with vector parameters $\lambda^{(0)}_i$ and $\lambda^{(1)}_i$ are to be taken into account,
and leads to constants of the motion
\bea
&&
\mathcal{C}^{(0)}_i=\rho^2 \dot\xi_i \left(\frac{\gamma^2}{\rho^4}+\frac{\dot\rho^4}{\gamma^2}-\frac{6 \dot\rho^2}{\rho^2}\right) +
16 \xi_i \frac{\dot\rho}{\rho} \left(\frac{\gamma^2}{\rho^2}-\dot\rho^2 \right),
\nonumber\\[2pt]
&&
\mathcal{C}^{(1)}_i=t \mathcal{C}^{(0)}_i+\rho^2 \dot\xi_i \left(\frac{3 \dot\rho}{\rho}
-\frac{\rho\dot\rho^3}{\gamma^2} \right)+4\xi_i \left(3\dot\rho^2-\frac{\gamma^2}{\rho^2} \right).
\eea
These are algebraic equations for $\xi_i$ and $\dot\xi_i$, which determine the evolution of $\xi_i$ with time.
Introducing a subsidiary function $s(t)$ as in (\ref{s}) and making use of (\ref{t}), one finds
\bea
\xi_i(t)=\alpha_i \cos{(4 \gamma s(t))}+\beta_i \sin{(4 \gamma s(t))},
\eea
where $\alpha_i$ and $\beta_i$ are constants of integration, which, if desirable, can be linked to
$\mathcal{C}^{(0)}_i$, $\mathcal{C}^{(1)}_i$, $\mathcal{H}$ and $\mathcal{D}$.

Thus, for $l=2$ there are two possibilities to realize the conformal Galilei algebra in terms of oscillator--like equations (\ref{L2})
with frequencies ${(2\gamma)}^2$ and ${(4\gamma)}^2$, respectively. As in the previous case, the corresponding orbits can be linked 
to those of the harmonic oscillators 
\bea
\frac{d^2}{ds^2} \chi_i +4 \gamma^2 \chi_i=0, \qquad
\frac{d^2}{ds^2}  \xi_i +16 \gamma^2 \xi_i=0,
\eea
by applying the map (\ref{s}).

\vspace{0.5cm}

\noindent
{\bf 4. The case of arbitrary $l$}\\

\noindent

Having considered two explicit examples, let us turn to the case of arbitrary $l$.
To begin with, let us rewrite the master equations (\ref{eqx}) in the matrix form
\be\label{ME}
\rho^2 \frac{d}{dt} x^{(n)}=x^{(m)} A^{mn},
\ee
where $x^{(n)}=(x^{(0)}, \dots, x^{(2l)})$ and $A^{mn}$ is a $(2l+1) \times (2l+1)$ matrix of the form
\be\label{matr}
A^{mn}=\left(
\begin{array}{ccccccc}
0 & -2 l \gamma^2 & 0 & 0 &\ldots & 0 & 0 \\
1 & 0 & -(2l-1) \gamma^2 & 0 &\ldots & 0 & 0 \\
0 & 2 & 0 & -(2l-2) \gamma^2 &\ldots  & 0 & 0 \\
0 & 0 & 3 & 0 &\ldots  & 0 & 0 \\
0 & 0 & 0 & 4 &\ldots  & 0 & 0 \\
\vdots &\vdots  & \vdots  & \vdots  &\ddots  & \vdots & \vdots \\
0 & 0 & 0 & 0 &\ldots  & -2\gamma^2 & 0 \\
0 & 0 & 0 & 0 &\ldots  & 0 & -\gamma^2 \\
0 & 0 & 0 & 0 &\ldots  & 2l & 0 \\
\end{array}
\right).
\ee
For the discussion to follow, the spatial index $i$ carried by $x^{(n)}$ is inessential and will be omitted.
The examples considered in the previous section indicate that the linear change of the fields in (\ref{chi}) and (\ref{nvar}) might have been related
to the eigenvectors of $A^{mn}$. Below we treat integer and half--integer values of $l$ separately as the two cases prove to be
qualitatively different.

For integer $l$ the matrix $A^{mn}$ is degenerate and has the following eigenvalues
\be
(0,\pm 2 i \gamma, \pm 4 i \gamma, \pm 6 i \gamma \dots, \pm 2l i \gamma).
\ee
Because $A^{mn}$ is real, all the eigenvectors occur in complex conjugate pairs, but for the eigenvector corresponding to the zero eigenvalue, which is real.
Let us denote the eigenvectors by $v^0_{(n)}$, $v^1_{(n)}$, ${\bar v}^1_{(n)}$, $\dots$,  $v^l_{(n)}$, ${\bar v}^l_{(n)}$, where the superscript refers to
the number of the corresponding eigenvalue, the bar stands for complex conjugate and $n=0,\dots,2l$. In particular, in this notation $v^1_{(n)}$ is related to the eigenvalue
$2i \gamma$, while ${\bar v}^1_{(n)}$ is linked to $-2i\gamma$. As usual, the eigenvectors are defined up to a factor.

Contracting the master equations (\ref{ME}) with the eigenvectors of $A^{mn}$ one gets
\bea\label{3}
&&
\rho^2 \frac{d}{dt} \left[ x^{(n)} v^0_{(n)} \right]=0,
\nonumber\\[2pt]
&&
\rho^2 \frac{d}{dt} \left[x^{(n)} (v^p_{(n)}+{\bar v}^{p}_{(n)})\right]=2 p \gamma \left[ i x^{(n)} (v^p_{(n)}-{\bar v}^{p}_{(n)})\right],
\nonumber\\[2pt]
&&
\rho^2 \frac{d}{dt} \left[i x^{(n)} (v^p_{(n)}-{\bar v}^{p}_{(n)})\right]=-2 p \gamma \left[ x^{(n)} (v^p_{(n)}+{\bar v}^{p}_{(n)})\right],
\eea
where $p=1,\dots,l$. Thus, it is natural to introduce the new fields
\be
x^{(n)} v^0_{(n)}, \qquad x^{(n)} (v^p_{(n)}+{\bar v}^{p}_{(n)}), \qquad i x^{(n)} (v^p_{(n)}-{\bar v}^{p}_{(n)}).
\ee
Because $x^{(n)} v^0_{(n)}$ obeys the first order equation, on physical grounds it seems reasonable to discard it.
The second line in (\ref{3}) allows one to express $i x^{(n)} (v^p_{(n)}-{\bar v}^{p}_{(n)})$ via $x^{(n)} (v^p_{(n)}+{\bar v}^{p}_{(n)})$.
The latter define a set of dynamical fields
\be\label{nf}
\chi_i^p=x^{(n)}_i (v^p_{(n)}+{\bar v}^{p}_{(n)}),
\ee
where $p=1,\dots,l$ and $i=1,\dots,d$, which
obey the equations of motion
\be\label{all}
\rho^2 \frac{d}{dt} \left(\rho^2 \frac{d}{dt} \chi^p_i \right)+{(2 \gamma p)}^2 \chi_i^p=0.
\ee
It is to be remembered that (\ref{all}) should be solved jointly with $\ddot\rho=\frac{\gamma^2}{\rho^3}$. In particular,
the $l=1$ and $l=2$ instances considered in the previous section are derived from
\begin{align}
&
v^0_{(n)}=\left(
\begin{array}{c}
\gamma\\[4pt]
0\\[4pt]
\frac{1}{\gamma} \\
\end{array}
\right), &&
v^1_{(n)}=\left(
\begin{array}{c}
\frac{\gamma}{2} \\[4pt]
-\frac{i}{2} \\[4pt]
-\frac{1}{2 \gamma} \\
\end{array}
\right), && {}
\end{align}
and
\begin{align}
&
v^0_{(n)}=\left(
\begin{array}{c}
3 \gamma^2 \\[4pt]
0 \\[4pt]
1 \\[4pt]
0 \\[4pt]
\frac{3}{\gamma^2} \\
\end{array}
\right), &&
v^1_{(n)}=\left(
\begin{array}{c}
\frac{\gamma^2}{2} \\[4pt]
-\frac{i\gamma}{4} \\[4pt]
0 \\[4pt]
-\frac{i}{4\gamma} \\[4pt]
-\frac{1}{2 \gamma^2} \\
\end{array}
\right), && v^2_{(n)}=\left(
\begin{array}{c}
\frac{\gamma^2}{2} \\[4pt]
-\frac{i\gamma}{2} \\[4pt]
- \frac 12\\[4pt]
\frac{i}{2\gamma} \\[4pt]
\frac{1}{2 \gamma^2} \\
\end{array}
\right),
\end{align}
respectively.

Note that, given $l$, (\ref{all}) contains a chain of oscillator--like equations with growing frequency. In particular, the value ${(2 \gamma )}^2$ appears for
any $l$, the equation involving ${(4 \gamma)}^2$ is common for all $l>1$, the frequency ${(6 \gamma)}^2$ is shared by all $l>2$ etc.
The reason why one can realize different $l$--conformal Galilei groups in one and the same equation is that all the vector generators $C^{(n)}_i$ with
$n>1$ prove to be functionally dependent on $C^{(0)}_i$ and $C^{(1)}_i$. To put it in other words, although $C^{(n)}_i$ with $n>1$ are involved in the
formal algebraic structure behind the equations of motion (\ref{all}),
they prove to be irrelevant for an actual solving thereof. If one had had a higher derivative formulation based on a
differential equation of the order $(2l+1)$, all $C^{(n)}_i$ would have been functionally independent and involved in the process of integration.
It is worthwhile drawing an analogy with the conformal mode
$\rho$ considered in Sec. 3. Special conformal transformations generated by $K$ are usually considered to be an attribute of the conformal mechanics.
However, $K$ is functionally dependent on $H$ and $D$ and constants of the motion corresponding to the time translation and the dilatation
are sufficient to integrate the equation of motion.

Now let us turn to a half--integer $l$. In this case the matrix (\ref{matr}) is nondegenerate and its eigenvalues read
\be
(\pm i\gamma, \pm 3 i \gamma, \pm 5 i \gamma, \dots, \pm 2 l i \gamma).
\ee
As before, the eigenvectors of $A^{mn}$ occur in complex conjugate pairs $v^{p}_{(n)}$, ${\bar v}^{p}_{(n)}$, where $p=1,3,5, \dots, 2l$, which prompt
one to introduce the new dynamical fields $\chi_i^p=x^{(n)}_i (v^p_{(n)}+{\bar v}^{p}_{(n)})$ and bring (\ref{ME}) to a set of decoupled generalized oscillators
\be\label{hi}
\rho^2 \frac{d}{dt} \left(\rho^2 \frac{d}{dt} \chi^{p}_i \right)+{(\gamma p)}^2 \chi_i^{p}=0.
\ee
As usual, (\ref{hi}) is accompanied by $\ddot\rho=\frac{\gamma^2}{\rho^3}$.
For example, for $l=\frac 32$ one finds the eigenvectors
\begin{align}
&
v^1_{(n)}=\left(
\begin{array}{c}
\frac{i \gamma^2}{2}\\[4pt]
\frac{\gamma}{6}\\[4pt]
\frac{i}{6} \\[4pt]
\frac{1}{2\gamma}
\end{array}
\right), &&
v^3_{(n)}=\left(
\begin{array}{c}
-\frac{i \gamma^2}{2}\\[4pt]
-\frac{\gamma}{2}\\[4pt]
\frac{i}{2} \\[4pt]
\frac{1}{2\gamma}
\end{array}
\right),
\end{align}
which determine the new fields
\be\label{exam}
\chi^1_i=\frac{1}{3} \gamma x^{(1)}_i+\frac{1}{\gamma} x^{(3)}_i, \qquad \chi^3_i=-\gamma x^{(1)}_i+\frac{1}{\gamma} x^{(3)}_i,
\ee
which obey the oscillator--like equations (\ref{hi}) with frequencies $\gamma^2$ and ${(3\gamma)}^2$, respectively.

As in the preceding case, (\ref{hi}) contains a chain of oscillator--like equations. In particular, those with frequencies $\gamma^2$ or
${(3\gamma)}^2$ may accommodate the $l$--conformal Galilei symmetry  with any half--integer $l$.

As was shown above, for integer and half--integer $l$ the matrix $A^{mn}$ has distinct properties. In the former case, $A^{mn}$ has an eigenvector
corresponding to the zero eigenvalue, which in essence indicates the resulting set of physical fields unambiguously. In the latter case, the choice of physical fields proves to be
ambiguous. This can be seen as follows. One can multiply (\ref{ME}) with the inverse of $A^{mn}$. This yields a chain of relations linking
$x^{(n)}_i$ to ${\dot x}^{(m)}_i$. Then one can try to remove any field. The only requirement to comply is that no higher derivative terms are produced
in the process. For example, within the scheme outlined above, the case $l=\frac 32$ is characterized by the equation (\ref{exam}), which implies
that $x^{(0)}_i$ and $x^{(2)}_i$ should be removed in favor of $x^{(1)}_i$ and $x^{(3)}_i$. A thorough investigation of other instances reveals one
more interesting possibility, which we discuss in detail in the next section.

\vspace{0.5cm}

\noindent
{\bf 5. Alternative realization of $l=\frac 32$}\\

\noindent

For $l=\frac 32$ the master equations (\ref{eqx}) read
\bea\label{l32}
&&
\rho^2 \dot x^{(0)}_i=x^{(1)}_i, \qquad \qquad \qquad \quad \rho^2 \dot x^{(3)}_i=-\gamma^2 x^{(2)}_i,
\nonumber\\[2pt]
&&
\rho^2 \dot x^{(1)}_i=2 x^{(2)}_i-3 \gamma^2 x^{(0)}_i, \qquad \rho^2 \dot x^{(2)}_i=3 x^{(3)}_i-2 \gamma^2 x^{(1)}_i.
\eea
Instead of introducing the physical fields as in (\ref{exam}), let us use
the first line in (\ref{l32}) to eliminate $x^{(1)}_i$ and $x^{(2)}_i$. Choosing the new variables $\phi_i$, $\psi_i$ as follows:
\be\label{ph}
\phi_i=\gamma x^{(0)}_i-\frac{1}{\gamma^2} x^{(3)}_i, \qquad \psi_i=\gamma x^{(0)}_i+\frac{1}{\gamma^2} x^{(3)}_i,
\ee
one brings the second line in (\ref{l32}) to the form\footnote{Note that the system of two second order differential equations (\ref{L32})
is equivalent one fourth order differential equation. Indeed, acting by the differential operator $\rho^2 \frac{d}{dt}$ on the first equation
in (\ref{L32}), one can relate $\rho^2 \frac{d}{dt} \left(\rho^2 \frac{d}{dt} \psi_i \right)$ to the derivatives of $\phi_i$. Substituting
this into the second line in (\ref{L32}) one can algebraically express $\psi_i$ via the derivatives of $\phi_i$. Inserting this $\psi_i$ back to
the first line in (\ref{L32}) one finds a fourth order differential equation for $\phi_i$. In order to treat (\ref{L32}) as a genuine two--body system it is
desirable to further deform it to include a nontrivial interaction potential compatible with $l=\frac 32$ conformal Galilei symmetry.}
\bea\label{L32}
&&
\rho^2 \frac{d}{dt} \left(\rho^2 \frac{d}{dt} \phi_i \right)+3 \gamma^2 \phi_i+2 \gamma \rho^2 \frac{d}{dt} \psi_i=0,
\nonumber\\[2pt]
&&
\rho^2 \frac{d}{dt} \left(\rho^2 \frac{d}{dt} \psi_i \right)+3 \gamma^2 \psi_i-2 \gamma \rho^2 \frac{d}{dt} \phi_i=0.
\eea
As usual, it is assumed that the field $\rho$ obeys the first equation in (\ref{L1}). Thus, the equation for $\rho$ in (\ref{L1})
along with (\ref{L32}) provide a dynamical realization of $l=\frac 32$ conformal Galilei
algebra, which is distinct from the decoupled formulation (\ref{hi}).

In order to solve
(\ref{L32}), one starts with the effective action
\be\label{S32}
S=\int dt \left( \rho^2 \dot\phi_i  \dot\phi_i+\rho^2 \dot\psi_i  \dot\psi_i  -3 \gamma^2 \frac{ \phi_i \phi_i}{\rho^2}
-3 \gamma^2 \frac{ \psi_i \psi_i}{\rho^2} -4\gamma \phi_i \dot\psi_i  \right),
\ee
where $\rho$ is a background field. Taking into account (\ref{ph}) and the second line in (\ref{transf}),
one then derives symmetries of (\ref{S32}), which correspond to the vector generators $C^{(n)}_i$ in $l=\frac 32$ conformal Galilei
algebra
\bea
&&
\delta \phi_i=\left(\frac{\gamma}{\rho^3}+\frac{\dot\rho^3}{\gamma^2} \right) \lambda^{(0)}_i+\left(t\left(\frac{\gamma}{\rho^3}+\frac{\dot\rho^3}{\gamma^2} \right)-
\frac{\rho \dot\rho^2}{\gamma^2} \right) \lambda^{(1)}_i+\left(t^2 \left(\frac{\gamma}{\rho^3}+\frac{\dot\rho^3}{\gamma^2} \right)
-2t \frac{\rho \dot\rho^2}{\gamma^2} +\frac{\rho^2 \dot\rho}{\gamma^2}\right) \lambda^{(2)}_i
\nonumber\\[2pt]
&&
\qquad \quad +\left(t^3 \left(\frac{\gamma}{\rho^3}+\frac{\dot\rho^3}{\gamma^2} \right)-3 t^2  \frac{\rho \dot\rho^2}{\gamma^2}+3t  \frac{\rho^2 \dot\rho}{\gamma^2}
-\frac{\rho^3}{\gamma^2}  \right) \lambda^{(3)}_i,
\nonumber\\[2pt]
&&
\delta \psi_i=\left(\frac{\gamma}{\rho^3}-\frac{\dot\rho^3}{\gamma^2} \right) \lambda^{(0)}_i+\left(t\left(\frac{\gamma}{\rho^3}-\frac{\dot\rho^3}{\gamma^2} \right)+
\frac{\rho \dot\rho^2}{\gamma^2} \right) \lambda^{(1)}_i+\left(t^2 \left(\frac{\gamma}{\rho^3}-\frac{\dot\rho^3}{\gamma^2} \right)
+2t \frac{\rho \dot\rho^2}{\gamma^2} -\frac{\rho^2 \dot\rho}{\gamma^2}\right) \lambda^{(2)}_i
\nonumber\\[2pt]
&&
\qquad \quad +\left(t^3 \left(\frac{\gamma}{\rho^3}-\frac{\dot\rho^3}{\gamma^2} \right)+3 t^2  \frac{\rho \dot\rho^2}{\gamma^2}-3t  \frac{\rho^2 \dot\rho}{\gamma^2}
+\frac{\rho^3}{\gamma^2}  \right) \lambda^{(3)}_i.
\eea
Noether theorem then yields four constants of the motion
\bea\label{com}
&&
{\mathcal{C}}^{(0)}_i=\left(\frac{\gamma}{\rho^3}+\frac{\dot\rho^3}{\gamma^2} \right) \rho^2 \dot\phi_i
+\left(\frac{\gamma}{\rho^3}-\frac{\dot\rho^3}{\gamma^2} \right) \rho^2 \dot\psi_i+
\left(\frac{2 \dot\rho^3}{\gamma}-\frac{3 \dot\rho^2}{\rho}+\frac{3 \gamma \dot\rho}{\rho^2}-\frac{2 \gamma^2}{\rho^3} \right) \phi_i+
\nonumber\\[2pt]
&& \qquad \quad
+
\left(\frac{2 \dot\rho^3}{\gamma}+\frac{3 \dot\rho^2}{\rho}+\frac{3 \gamma \dot\rho}{\rho^2}+\frac{2 \gamma^2}{\rho^3} \right) \psi_i,
\nonumber\\[2pt]
&&
{\mathcal{C}}^{(1)}_i=t {\mathcal{C}}^{(0)}_i-\frac{1}{\gamma^2} \rho^3 \dot\rho^2 \dot\phi_i+\frac{1}{\gamma^2} \rho^3 \dot\rho^2 \dot\psi_i-
\left(\frac{\gamma}{\rho}-2\dot\rho+\frac{2\rho\dot\rho^2}{\gamma} \right)\phi_i-
\left(\frac{\gamma}{\rho}+2\dot\rho+\frac{2\rho\dot\rho^2}{\gamma} \right)\psi_i,
\nonumber\\[2pt]
&&
{\mathcal{C}}^{(2)}_i=t {\mathcal{C}}^{(1)}_i+\left( \frac{\rho^2 \dot\rho}{\gamma^2}-t \frac{\rho \dot\rho^2}{\gamma^2}\right)\rho^2 \dot\phi_i
-\left( \frac{\rho^2 \dot\rho}{\gamma^2}-t \frac{\rho \dot\rho^2}{\gamma^2}\right)\rho^2 \dot\psi_i
-\left(t\left(\frac{\gamma}{\rho}-2 \dot\rho+\frac{2\rho \dot\rho^2}{\gamma} \right)-
\right.
\nonumber\\[2pt]
&& \qquad \quad
\left.
 -\frac{2 \rho^2 \dot\rho}{\gamma}+\rho\right)\phi_i
-\left(t\left(\frac{\gamma}{\rho}+2 \dot\rho+\frac{2\rho \dot\rho^2}{\gamma} \right) -\frac{2 \rho^2 \dot\rho}{\gamma}-\rho\right)\psi_i,
\nonumber\\[2pt]
&&
{\mathcal{C}}^{(3)}_i=t {\mathcal{C}}^{(2)}_i-\left(t^2 \frac{\rho \dot\rho^2}{\gamma^2}-2t\frac{\rho^2 \dot\rho}{\gamma^2}+\frac{\rho^3}{\gamma^2}\right)\rho^2\dot\phi_i
+
\left(t^2 \frac{\rho \dot\rho^2}{\gamma^2}-2t\frac{\rho^2 \dot\rho}{\gamma^2}+\frac{\rho^3}{\gamma^2}\right)\rho^2\dot\psi_i-
\nonumber\\[2pt]
&& \qquad \quad
-\left(t^2\left(\frac{\gamma}{\rho}-2\dot\rho+\frac{2\rho\dot\rho^2}{\gamma}\right)-2t \left(\frac{2\rho^2\dot\rho}{\gamma}-\rho \right)+\frac{2\rho^3}{\gamma}\right)\phi_i-
\nonumber\\[2pt]
&& \qquad \quad
-\left(t^2\left(\frac{\gamma}{\rho}+2\dot\rho+\frac{2\rho\dot\rho^2}{\gamma}\right)-2t \left(\frac{2\rho^2\dot\rho}{\gamma}+\rho \right)
+\frac{2\rho^3}{\gamma}\right)\psi_i,
\eea
which can be viewed as a system of linear inhomogeneous algebraic equations for $\phi_i$, $\psi_i$, $\dot\phi_i$, $\dot\psi_i$.
This yields the general solution to the equations of motion (\ref{L32}).
Most conveniently it is written in terms of the subsidiary function (\ref{s}) with the use of the dictionary (\ref{t})
\bea\label{LL32}
&&
\phi_i(t)=\alpha_i \cos{(\gamma s(t))}+\beta_i \sin{(\gamma s(t))}+\mu_i \cos{(\gamma s(t))} \sin^2{(\gamma s(t))}+
\nonumber\\[2pt]
&&
\qquad \quad +\nu_i \sin{(\gamma s(t))} \cos^2{(\gamma s(t))}=
\left(\alpha_i+ \beta_i \frac{(\mathcal{D}+t \mathcal{H})}{\gamma} \right) {\left[1+\frac{{(\mathcal{D}+t \mathcal{H})}^2}{\gamma^2} \right]}^{-\frac{1}{2}}+
\nonumber\\[2pt]
&&
\qquad \quad +\left(\nu_i+\mu_i \frac{(\mathcal{D}+t \mathcal{H})}{\gamma}\right) \frac{(\mathcal{D}+t \mathcal{H})}{\gamma}
{\left[1+\frac{{(\mathcal{D}+t \mathcal{H})}^2}{\gamma^2}\right]}^{-\frac{3}{2}},
\nonumber\\[2pt]
&&
\psi_i(t)=\beta_i \cos{(\gamma s(t))}-\alpha_i \sin{(\gamma s(t))}+\nu_i \cos{(\gamma s(t))} \sin^2{(\gamma s(t))}
\nonumber\\[2pt]
&& \qquad \quad
-\mu_i \sin{(\gamma s(t))}
\cos^2{(\gamma s(t))}=
\left(\beta_i-\alpha_i \frac{(\mathcal{D}+t \mathcal{H})}{\gamma} \right) {\left[1+\frac{{(\mathcal{D}+t \mathcal{H})}^2}
{\gamma^2} \right]}^{-\frac{1}{2}}-
\nonumber\\[2pt]
&&
\qquad \quad -\left(\mu_i-\nu_i \frac{(\mathcal{D}+t \mathcal{H})}{\gamma}\right) \frac{(\mathcal{D}+t \mathcal{H})}{\gamma}
{\left[1+\frac{{(\mathcal{D}+t \mathcal{H})}^2}{\gamma^2}\right]}^{-\frac{3}{2}}.
\eea
Here $\alpha_i$, $\beta_i$, $\mu_i$ and $\nu_i$ are constants of integration, which, if desirable, can be expressed in terms of
${\mathcal{C}}^{(0)}_i$, ${\mathcal{C}}^{(1)}_i$, ${\mathcal{C}}^{(2)}_i$, ${\mathcal{C}}^{(3)}_i$, $\mathcal{H}$ and $\mathcal{D}$.

To summarize, the second order differential equations (\ref{L32}) along with the first equation in (\ref{L1}) provide a dynamical realization
of $l=\frac 32$ conformal Galilei algebra.  As compared to the previous case, all the four vector generators $C^{(n)}_i$ prove to be functionally
independent, which correlates with the fact that one needs four independent constants of the motion in order to integrate
two second order differential equations.

As above, a relation between $t$ and $s$ in (\ref{s}) can be used to link (\ref{L32}) to a system of two coupled oscillators in $d$ dimensions
\bea\label{O32}
&&
\frac{d^2 \phi_i}{ds^2}  +3 \gamma^2 \phi_i+2 \gamma \frac{d \psi_i}{d s} =0, \qquad
\frac{d^2 \psi_i}{ds^2}  +3 \gamma^2 \psi_i-2 \gamma \frac{d \phi_i}{d s} =0.
\eea
In view of the considerations above and taking into account the analysis in \cite{gm},
this system is likely to possess $l=\frac 32$ conformal Newton-Hooke symmetry. This issue will be studied elsewhere.

\vspace{0.5cm}

\noindent
{\bf 6. Conclusion}\\

\noindent

To summarize, in this work the method of nonlinear realizations was applied to
the $l$--conformal Galilei algebra to construct a dynamical system
without higher derivative terms in the equations of motion.
A configuration space of the model involves coordinates, which
parametrize particles in $d$ spatial dimensions and
a conformal mode, which gives rise to an effective external field.
Given $l$, the dynamical realization involves a set of decoupled
oscillator--like equations each of which holds invariant under the $l$--conformal Galilei group.
A map of the orbits traced by the particles into those of
a set of decoupled oscillators in $d$ dimensions was constructed.
The status of the acceleration generators within the scheme was shown to be analogous to that of the generator of special conformal transformations in the conformal
mechanics. Although accelerations are involved in the
formal algebraic structure behind the equations of motion,
they prove to be functionally dependent. 

Turning to possible further developments, the most urgent question is whether the dynamical equations constructed in this work
can be deformed so as to include interaction terms in a way compatible with the $l$--conformal Galilean symmetry. This question is also applicable to the
formulation in Sect. 5. The dynamical realization we proposed
in this work is not Lagrangian. It would be interesting to find a set of auxiliary fields leading to a Lagrangian formulation.
A generalization of the analysis in Sect. 5 to the case of arbitrary half--integer $l$ is also an interesting open problem.

\vspace{0.5cm}

\noindent{\bf Acknowledgements}\\

This work was supported by the Dynasty Foundation, RFBR grant 12-02-00121, LSS
grant 224.2012.2, MSE grant "Nauka" 1.604.2011, and RF Federal Program "Kadry" under the project 2012-1.2.1-12-000-1012-4521.

\end{document}